\documentclass[a4paper,12pt]{article}

\usepackage{citesort}
\usepackage{amsmath}
\usepackage{graphicx}

\def\beq{\begin{equation}}
\def\be{\begin{equation}}
\def\eeq{\end{equation}}
\def\ee{\end{equation}}
\def\bea{\begin{eqnarray}}
\def\eea{\end{eqnarray}}

\def\bq{\begin{quote}}
\def\eq{\end{quote}}

\def\gappeq{\mathrel{\rlap {\raise.5ex\hbox{$>$}}
{\lower.5ex\hbox{$\sim$}}}}
\def\lappeq{\mathrel{\rlap{\raise.5ex\hbox{$<$}}
{\lower.5ex\hbox{$\sim$}}}}
\def\Toprel#1\over#2{\mathrel{\mathop{#2}\limits^{#1}}}

\allowdisplaybreaks

\begin{document}

\begin{titlepage}
\vspace*{-2cm}

{\Large
\begin{center}
{\bf Minimal Prescription Corrected Spectra \\ in Heavy Quark
Decays}
\end{center}
}
\vspace{.5cm}

\begin{center}
{L.~Di~Giustino~$^{a}$,~~~~G.~Ricciardi~$^{b}$,~~~~L.~Trentadue~$^{c}$}
\\[7mm]
{$^{a}$\textit{SLAC National Accelerator Laboratory, \\
Stanford University, Stanford, CA 94309, USA}}\\
[4mm]
{$^{b}$\textit{Dipartimento di Fisica, Universit\`a di
Napoli "Federico II"
\\
and
\\
INFN Sezione di Napoli, Napoli, Italy }}\\
[4mm] {$^{c}$\textit{Dipartimento di Fisica, Universit\`a di
Parma,
\\
and
\\
INFN, Gruppo Collegato di Parma, Parma, Italy}}\\
[10pt]   \vspace{1.5cm}

\begin{abstract}


The Minimal Prescription procedure is applied to tame the Landau pole singularities of resummed formulae for heavy quark decays. Effects of the final quark mass are taken into account.
Explicit expressions  are obtained for the $t \rightarrow b$ and $b \rightarrow c$  transitions for both the frozen coupling approximation and in the QCD running coupling case.
\end{abstract}

\end{center}

\end{titlepage}

\noindent

\section{Introduction}

A common feature to many processes in QCD is the presence, in the perturbative expansion, of large double (Sudakov-like) logarithms at the threshold.
Resummation of large infrared logarithms in form factors and shape
variables is essential in order to predict accurate cross sections
in many phenomenologically relevant processes
(see, f.i. \cite{parpet,kodtren,sterman,cattren,cattren2}).
In semi-leptonic heavy quark decays $ q_i \rightarrow q_f \, l \, \nu$, threshold regions are characterized by the presence of two different scales $ m_X \ll E_X$, where $m_X$ and $E_X$ are  the final hadron invariant mass and energy, respectively, originated by the final quark $q_f$. The perturbative expansion is spoiled by logarithms of the ratio of the two scales. Those need, therefore, to be resummed.

\noindent Such logarithms are organized as a series of the
form \cite{aglietti:2001,Aglietti:2005mb}:
\begin{eqnarray}
\label{finallargelogs}
&&
\sum_{n=1}^{\infty}
\sum_{k=1}^{2n}
c_{n k} \,\alpha_s^n(Q) \, \log^k \frac{Q^2}{m_X^2}
=\;\;\; c_{12}\,\alpha_s(Q)\,\log^2\frac{Q^2}{m_X^2}
\, + \, c_{11} \, \alpha_s(Q)\,\log\frac{Q^2}{m_X^2}
\, +\\
&+& \, c_{24} \, \alpha_s^2(Q)\,\log^4\frac{Q^2}{m_X^2}
\, + \, c_{23} \, \alpha_s^2(Q)\,\log^3\frac{Q^2}{m_X^2} \, + \, \cdots,\nonumber
\end{eqnarray}
where $\alpha_s$ is the QCD coupling constant and
$Q $ is the hard scale  $Q= 2\, E_X$. The  leading term is the double logarithm  $\alpha_s(Q)\,\log^2\frac{Q^2}{m_X^2}$.
A similar double logarithmic structure is present in many other processes like deep inelastic scattering (DIS), heavy quark fragmentation, Drell-Yan annihilation, Higgs
production,  and so on, the  argument of the  logarithms differing per observable and per process.

\noindent A universal resummation formula valid at all perturbative orders for a decay of a heavy quark $q_i$ into a massive quark $q_f$ plus a non-hadronic state,  with a final state jet-like structure, has been recently obtained \cite{nostro2007-1}.

In QCD resummed formulas,  the
running coupling is integrated over
all gluon radiative momenta  from the hard scale down to
zero, hitting the Landau pole. A
prescription has to be assigned to give a meaning to the formal
resummed expressions.

One possible solution is the use of an additional prescription for the contour
integration in $N$-space, in the inverse Mellin transform from $N$-space to
$x$-space, the so-called minimal-prescription (MP)
\cite{Catani:1996}.
This prescription provides a formula
which is the asymptotic limit of the expansion, furthermore it is
renormalon free and the truncation of the series at the minimum
term originates an exponentially suppressed difference between the
truncated expansion and the full MP formula.

The aim of this work is to analyze the perturbative resummed distributions in the parton subprocess for both massless and massive final heavy quarks and to explore the feasibility of the MP regularization scheme.

We apply the resummation formulas to
the case of  $t \rightarrow b$ and $b \rightarrow c$, as a working example to implement this regulation method.
In literature, the  $b \rightarrow c$ inclusive semileptonic decays are widely discussed,  also in the contest of effective theories, in order to improve comparison with the newest data (for a review see f.i \cite{Gambino:2011fz} and refs within).
The  $t \rightarrow b$ case has  been  discussed in QCD resummed formulas \cite{Cacciari:2002re}, with  different dynamical variables.

The assessment of a perturbative reliable and singularity safe form factor is the first step towards a sound phenomenological approach and  it is also  needed  for comparison with
 QCD based effective theories like, for example, SCET.

The papers is organized as follows:
in section \ref{Mellin} we recall the resumming formulas for the final massless and massive quark in the Mellin space, and recast them in a more transparent notation; in section \ref{subsec41} we move to the physical space and analyze the feasibility of the MP for the massive case. In section \ref{FC-section} we study the frozen coupling approximation, while final plots with the QCD running coupling and conclusions are presented in sections \ref{conclusionsec} and \ref{conclusionfin}.

\section{Threshold resummed Jet Distribution in Mellin space}
\label{Mellin}

\subsection{Massless final quark}
\label{subsec3}

Before considering the case of a massive final case,
let us recall the expressions and the variables  for the  resummed jet distributions in the massless final state \cite{sterman,catanitrentadue,Catani:1990rr,catcac,sghedoni}.

Let us consider the decay driven at a partonic level by an heavy quark decaying into an approximately massless final quark, plus non-hadronic states, as, for instance,  the decay $B \rightarrow X_u l \nu$ or the radiative decay $B \rightarrow X_s \gamma$.
Threshold resummation is typically performed in  Mellin space;
 the threshold limit corresponds to $N \rightarrow \infty$ and threshold logarithms $ \alpha_s^n \log^m N$ can be factorized into a
 form factor $J_N$, which  has the exponential form: \beq \label{generale} J_N(Q^2) \, = \, e^{\, f_N \left(Q^2 \right) }, \eeq
 $J_N(Q^2)$ is the
massless jet distribution,  that gives the probability that a massless parton produced in a hard process with a  hard scale  $Q$ fragments into a hadronic jet of mass $m_X$
\beq
m^2_X = (1-x) Q^2
\label{defx}
\eeq
The Mellin or $N-$ transform   is defined as
\beq
J_N(Q^2)  \equiv   \int_0^1 dx \, x^{N-1}  J(x; Q^2)
\label{Mellin-def}
\eeq

In the limit $\alpha_s \rightarrow 0$, the mass distribution reduces to a spike
corresponding to the (zero) parton mass. In the limit $x\rightarrow 1$ we drift away from the perturbative regime.
If $x =1$, the
 truncated perturbative expansion becomes unreliable. It is possible, however, to be able to use a  perturbative
 resummed expression at all orders in the Mellin space which reads \cite{sterman,catanitrentadue,Catani:1990rr,catcac,sghedoni}:
 \bea \label{gen-resumming1} J_N(Q^2) &=& \exp \int_0^1 \frac{dx}{1-x} \left[ x^{N-1} -
1 \right]
 \Bigg\{ \int_{ Q^2 (1-x)^2 }^{ Q^2 (1-x) }
\frac{dk_{\perp}^2}{k_{\perp}^2}
A\left[\alpha_s\left(k_{\perp}^2\right)\right] \, + \, B
\left[\alpha_s\left(Q^2 (1-x)\right)\right] \, + \,
\nonumber\\&&+D \left[\alpha_s\left(Q^2 (1-x)^2\right)\right]
 \Bigg\} \eea
The functions $A\left(\alpha_s\right)$, $B\left(\alpha_s\right)$ and
$D\left(\alpha_s\right)$ have a perturbative expansion:
\begin{equation}
 A\left(\alpha_s\right) =
  A_{1}\alpha_s+A_{2}\alpha_s^{2}+\cdots, \quad
B\left( \alpha_s\right) = B_1\alpha_s+B_2\alpha_s^2+\cdots,
\quad D\left(\alpha_s\right) =D_1\alpha_s+D_2\alpha_s^2+\cdots .
\end{equation}
The known values of the coefficients $A_i$, $B_i$ and $D_i$ are given in \cite{Catani:1990rr,eroi,kt}.

$A\left(\alpha_s\right)$ describes the emission of partons which
are both soft and collinear, $B\left(\alpha_s\right)$ describes
hard and collinear partons while $D\left(\alpha_s\right)$
partons which are emitted soft at large angles.
$A(\alpha_s)$ and $B(\alpha_s)$ are related to small-angle emission
only. They, therefore, represent intra-jet properties\cite{kt,eroi}, while
the function $D(\alpha_s)$, being related to soft emissions at large
angles, is a process-dependent inter-jet quantity.

While the validity of the resummed formula  goes beyond our case of  semi-leptonic heavy quark decays,
holding for accounting of threshold logarithms in several other processes, the specific structure
of (\ref{gen-resumming1}) can vary, depending on the specific process and on the particular observable under exam.
For instance,  there are corresponding results for the DIS structure
functions $F_{1,2,3}(x,Q^2)$, where $Q^2$ represents the resolution scale,
or for the Drell-Yan cross section $d\sigma/dQ^2$, where $Q^2$ stands for the invariant mass squared of the lepton pair \cite{catanitrentadue,vogt:00}.

In order to illustrate how to interpret the universality of formula (\ref{gen-resumming1}), let us consider the order $\alpha_s$ decay $ t \rightarrow \, b\, W \, g$ (where
$W$ and $g$ are a real  W boson and a gluon, respectively) and examine
the  distribution in the energy of the final $b$-quark, that is in the variable $x_b= 2 E_b/m_t$.
Once considering the distribution in $x_b$, rather than in $x$ as in the present paper (see definition (\ref{defx})),  we are dealing with  a different
 observable--and therefore a different kinematical parametrization of the threshold region.
Now the threshold
region is reached when $x_b \rightarrow 1 $, a limiting point where  there is no gluon emission to change the light quark energy.
We expect only the emission of soft gluons and we do not need  any $B(\alpha_s)$  contribution in (\ref{gen-resumming1}), since this function contains collinear radiation associated with the light quark. Formula (\ref{gen-resumming1}) still holds, but without the
$B(\alpha_s)$ term\cite{Cacciari:2002re}.
In the present work,  collinear gluons described by $B(\alpha_s)$ are allowed, since  the energy change of the light quark,  still  in the jet after the gluon emission, does not affect the distributions in $x$, related to the invariant mass of the jet.

The exponent of Eq.~(\ref{generale}) can be expanded in a function series of the form
\cite{catanitrentadue}:
\begin{equation}
\label{series}
f_N (Q^2)= \log J_N  (Q^2)=
L\,g_1\left(\lambda\right) +\sum_{n=0}^{\infty
}\alpha_s^n\,g_{n+2}\left(\lambda\right)  =
L\,g_1\left(\lambda\right) +\,g_2\left(\lambda\right)
+\alpha_s\,g_3\left(\lambda\right) +\cdots,
\end{equation}
where
\begin{equation}
\lambda~=~\beta_0~\alpha_s(Q^2)~L,~~~~~~~L~=~\log N
\end{equation}
and $\beta_0 = (11/3\;N_C-2/3\; n_F)/(4\pi)$.

The first exponential term
$
L\,g_1\left(\lambda\right) =
L\,\sum_{n=1}^{\infty }g_{1,n}\lambda^n
$ resums the leading logarithms (LL); by adding
the term $
g_2\left(\lambda\right) =
\sum_{n=1}^{\infty }g_{2,n}\lambda^n
$, also next-to-leading order terms (NLL) are taken into account and resummed, and so on.

 The functions
$g_i\left( \lambda \right)$
have a power-series expansion:
\begin{equation}
g_i\left(\lambda \right) =\sum_{n=1}^{\infty }g_{i,n}\lambda^n.
\end{equation}
They are all homogeneous functions: $g_i (0) = 0$. This property insures the normalization
of the form factor $J_{N=1} = 1$.
The functions $g_{1}$ and $g_{2}$ become singular, signaling non-perturbative effects, at
$\lambda=1/2$, that is at
$N= \exp[1/2 \beta_0 \alpha_s(\mu^2)] \approx \mu^2/\Lambda^2 $.
Explicit expressions  are given in \cite{Aglietti:2002ew,americani}.

The leading logarithmic term in Eq.~(\ref{gen-resumming1}) is \beq J_N \simeq \exp \left[ L\,g_1\left(\lambda\right) \right]
\simeq \exp \left[ - \frac{A_1}{2} \alpha_s L^2\right] = \exp \left[ - \frac{C_F}{2 \pi} \alpha_s \log^2 N\right]\eeq
Function series like Eq.~(\ref{series})  appear in other processes as well as, for instance, DIS and Drell-Yan with the same towers of threshold logarithms. Additional terms, however, due to soft-gluon radiation collinear to the  light initial-state parton, in the DIS case, and to two light initial-state partons, in the Drell-Yan case, do slightly modify the form of $g_1$, giving, for the leading term of the resummed quark coefficient
functions $C^N_{DIS}$ and $C^N_{DY}$\cite{vogt:00} respectively :
\beq C^N_{DIS} \propto \exp \left[ L\,g_1^{DIS}\left(\lambda\right) \right]
\simeq \exp \left[  \frac{C_F}{2 \pi} \alpha_s \log^2 N\right]\qquad
C^N_{DY} \propto \exp \left[ L\,g_1^{DY}\left(\lambda\right) \right]
\simeq \exp \left[  2 \frac{C_F}{ \pi} \,\alpha_s \log^2 N\right]\eeq

We plot in fig. 1 the form factor in the massless case, for the $ t \rightarrow b$  and the $ b \rightarrow$ decays, in order to show the different slopes in the two cases, that will be maintained in the massive case, affecting the regularization procedure.
We have set the scale $\mu$ in $g_i$ equal to $Q^2$.
The continuous and dashed lines represent  the NLL  and NNLL contributions respectively. The strong dependence on $\alpha_s$ values  is shown: light and thick lines are referred to different values of $\alpha_s$. The NNLL curve stands below the NLL one, due to the inclusion of $g_3$, that is a negative decreasing function within the considered range.
We also observe a relatively strong dependence on $\alpha_s$. As said before, the effects of the Landau pole start appearing at  $\lambda=1/2$, that is towards  $N \sim 10^6$ for the case of the top decay, and for $N \sim 10^2$ for the case of bottom decay;  there, the Mellin form factor starts to oscillate and the expressions are no longer predictive.

\begin{figure}[htb]
\includegraphics[width=21pc]{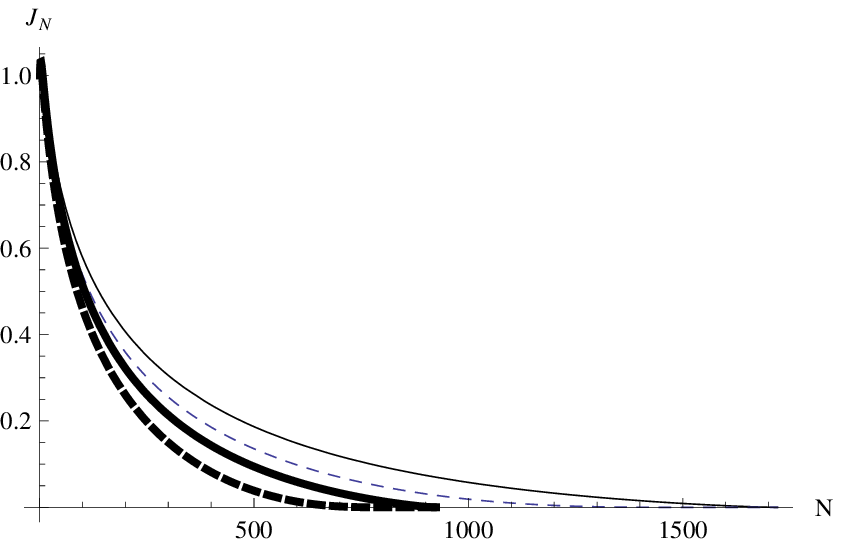}
\includegraphics[width=21pc]{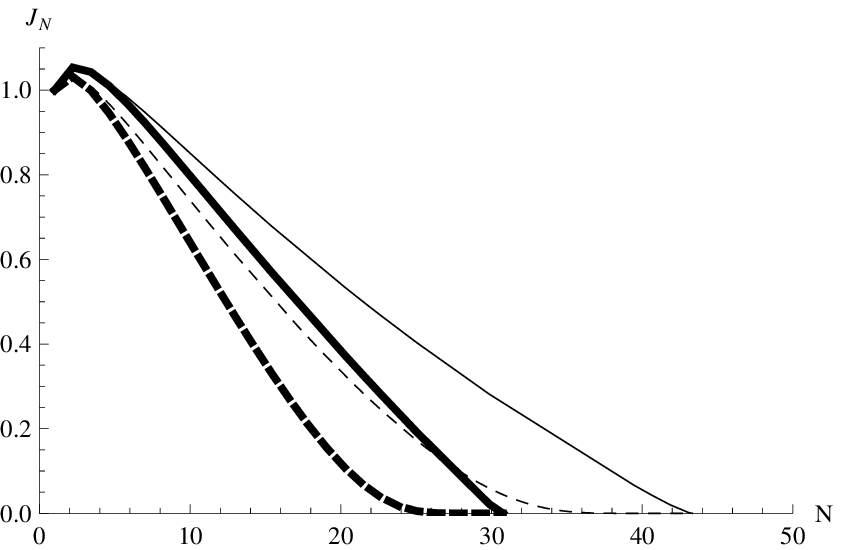}
\caption{Form factor in N space for massless final quark. Left figure:
 the $t \rightarrow b$ case: $\alpha_s= 0.11$ (light lines), $\alpha_s= 0.12$ (thick lines).  Right figure: the $b \rightarrow c$ case, $\alpha_s= 0.20$ (light lines), $\alpha_s= 0.22$  (thick lines).
 In both figures the continuous lines  represent  NLL contributions,  the dashed lines  NNLL contributions.} \label{form-fact-4}
\end{figure}

\subsection{Mass-corrected Jet Distribution}
\label{subsecMass1}

Let us briefly summarize the results obtained for the case of a massive final quark $q_f$ \cite{nostro2007-1}. The more massive is the radiating $q_f$, the less radiation has to be emitted in the decay; as a consequence, the typical Sudakov effect, namely the suppression of non-radiative
channels and the broadening of sharp structures, are expected to be less pronounced for the massive channels.
In principle one has single-logarithmic corrections, which are not strong enough to shift the peak of tree-level distributions.

In  Ref.\cite{nostro2007-1} it was demonstrated at NNL order (and conjectured to be also valid at accuracy beyond NNL) that the massive jet function can be factorized in momentum space,
 as: \beq \label{univ} J_N(Q^2; \,r)
\, = \, J_N(Q^2) ~ \delta_N(Q^2; \,r) \eeq $J_N(Q^2)$ is the
massless jet distribution and
$\delta_N(Q^2; \,r) $ is the mass-correction factor which
reads: \bea \label{maineq}  \delta_N(Q^2; \,r)  &=& \exp \int_0^1 d
x \frac{ x^{ \, r \, (N-1)} - 1 }{1-x} \Bigg\{
 - \int_{ m^2 (1-x)^2 }^{ m^2 (1-x) } \frac{dk_{\perp}^2}{k_{\perp}^2} A\left[\alpha_s\left(k_{\perp}^2\right)\right]
 - B \left[\alpha_s\left( m^2 (1-x) \right)\right]
+ \nonumber\\
&&\,
+ \, D \left[\alpha_s\left( m^2 (1-x)^2 \right)\right] \Bigg\}, \eea
 $x$  and $r$ are defined as \beq \label{eqdefy} y \equiv 1-x \, \equiv \, \frac{m_X^2 -
m^2}{Q^2 - m^2} \qquad  \qquad r \,
\equiv \, \frac{m^2}{Q^2} \, \ll \, 1\eeq
where  $Q$ is the hard scale of the process and  $m$ is
the mass of the emitting quark.
We  assume the quark mass to be much smaller than the
hard scale,  in order to have fast-moving charges and to preserve a jet
structure.
We indicate both the mass corrected and the massless jet functions with $J_N$; they are   distinguishable since the
 the massive one bears a dependence on  $r$.

Eq.~(\ref{maineq}) has a simple physical
interpretation. The parameter $N-1$ is multiplied by $r$ on the
r.h.s. of Eq.~(\ref{maineq}), implying that mass effects become
``visible'' only for large \beq N \, \geq \, \frac{1}{r} \, \gg \,
1. \eeq In this case, there is enough resolution to ``see'' the
quark mass, which tends to suppress the collinear effects, related
to the $A$ and $B$ terms. At the same time, soft radiation not
collinearly enhanced, described by the function $D$ and
characteristic of massive partons, does appear. Let us also note
that, since the jet mass is an infrared (i.e. soft and collinear)
safe quantity, $\delta_N = 1$ for $r = 0$. In the limit $r \, \to
\, 0$, the well-known massless result is recovered.

 The mass-correction factor has the same
structure than the massless case
\cite{catanitrentadue} \beq \delta_N(Q^2; \,r) \, =
\, e^{ F_N \left( Q^2;r \right) }, \eeq where the exponent has a
double expansion of the form: \beq F_N\left( Q^2; r \right) \, =
\, \theta\left( N - 1/r \right)\, \sum_{n=1}^{\infty}
\sum_{k=1}^{n+1} F_{n k} \, \alpha_s^n \log^k (N r), \eeq with $F_{n
k}$ numerical coefficients. The exponent can be expanded in towers
of logarithms as: \bea F_N\left( Q^2; r \right) &=& L \, d_1
\left( \rho \right) \, + \, \sum_{n=0}^{\infty} \alpha_s^n \,
d_{n+2}\left( \rho \right)
\nonumber\\
&=& L \, d_1 \left( \rho \right) \, + \, d_2 \left( \rho \right)
\, + \, \alpha_s \, d_3\left( \rho \right) \, + \, \alpha_s^2 \,
d_4\left( \rho \right) \, + \, \cdots, \eea where \beq \rho \,
\equiv \, \beta_0 \alpha_s(\mu^2) \, L ,\,\, \mathrm{and} \,\,\,
 L =\, \, \theta\left( N - 1/r \right) \, \log \left( N \, r
 \right) .\eeq The scale $\mu = O(m)$ is a renormalization scale
 of the order of the quark mass $m$. The over-all factor
$\theta\left( N - 1/r \right)$ comes from the step approximation
of the moment kernel and avoids modifications for small $N$ of the
massless behavior, in agreement with the physical intuition. Furthermore
it ensures the correct massless behavior in the $r \rightarrow 0$
limit. Analytic continuation to the complex $N$-plane can be made
by omitting such factor and fixing the correct interval in physical space.

\noindent By truncating the above series expansion, one obtains a
fixed-logarithmic approximation to the form factor $\delta_N$.
Functions $d_i(\rho)$, which represent the mass effects, can be
obtained from the standard ones $g_i(\lambda)$ of the massless
case \cite{Aglietti:2005mb} by means of the replacements:
 \beq
A(\alpha_s) \, \to \, - \, A(\alpha_s); ~~ B(\alpha_s) \, \to \, - \,
B(\alpha_s); ~~ D(\alpha_s) \, \to \, D(\alpha_s); ~~
\log\frac{\mu^2}{Q^2} \, \to \, \log\frac{\mu^2}{m^2} ; ~~ \lambda
\, \to \, \rho. \eeq
It is worth observing that mass effects
induce a similar structure to the massless one, involving changes
of sign of the collinear functions $A$ and $B$, with the rescaling
$Q \, \to \, m$.
The explicit expressions for the functions $d_i$ are listed in \cite{nostro2007-1}.

Let us now examine the behaviour of the jet function as given by the equation (\ref{univ}).
Throughout the paper we fix the hard scale of the process $Q$ to the mass of the decaying
quark, that is to $m_t$ in the case of top decays, and to
$m_b$ for b decays.
The correction factor $\delta_N(Q^2; \,r)$ is a function increasing with $N$; in order to produce $J_N(Q^2; \,r)$ it has to be multiplied by the massless form factor  $J_N(Q^2)$, at values  of $ N > 1/r$. In the case of the top quark decay, the increase is very slow; it starts at  $1/r \sim 2 \, \textrm{x} \, 10^3$ and the distribution only doubles when $N$ reaches $\sim 6 \, \textrm{x} \, 10^4$, continuing slowly, until a fast increase before values of $\sim 2 \, \textrm{x}\, 10 \, \, m_t^2/\Lambda^2 \sim 10^6$, where it reaches the peak \footnote{Differences between the mass correction factor  at NNLL and NLL order are that the NNLL corrected one peaks much faster and at an earlier point in $N$;  the increase, however, still occurs at the same order of magnitude and it does not introduce substantial changes on the distribution.}.
That implies, as expected, that mass addition does not modify substantially the massless distribution until very large values of $N$.  The left hand plot in fig. 1 practically coincides with  $J_N(Q^2, r)$ defined in (\ref{univ}),  at the same value of $\alpha_s$.

\noindent On the contrary, in the $b \rightarrow c$ case, the increase starts earlier ($1/r \sim 2 \, \textrm{x} \, 10$) and it  is much faster (it doubles one at $N \sim 10^2$), reaching the fast increase and the peak around  $N \sim m_b^2/\Lambda^2$;  effects are much more sizable.
NNLL corrected curves grow faster  than the NLL curves, as shown in fig. 2.

\begin{figure}[htb]
\begin{center}
\includegraphics[width=21pc]{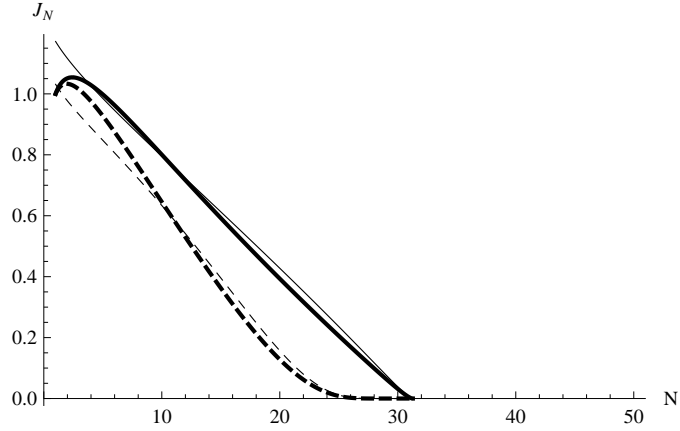}
\caption{Form factor in N space for $b \to c$ at $\alpha_s=0.219$: the continuous and dashed lines represent  NLL  and NNLL contributions, respectively.
Light and  thick lines refer here to massive and massless final quark, respectively.}
\end{center} \label{form-fact-4}
\end{figure}

\section{Threshold resummed Jet Distribution in physical space }
\label{subsec41}

Even if the $N$-moment expressions of the jet function are physical quantities, their measurement, especially for large $N$, is difficult. It is therefore convenient to perform
the inverse Mellin transform back to momentum space.
Given the Mellin transform    $J_N$ defined as in (\ref{Mellin-def}), its inverse transform is
\beq
J(x; Q^2) = M^{-1}[J_N; x]= \frac{1}{2 \pi i} \int_{C -i \infty}^{C + i\infty} dN \, x^{-N} J_N (Q^2)
\label{inv-m2}
\eeq

The inverse transform of the product of two generic $f_N$ and $g_N$ is the  convolution of the  two inverse functions $f(x)$ and $g(x)$:
\beq
M^{-1}[f_N \, g_N; x]=\int_x^1 f\left(\frac{x}{u}\right) g(u) \frac{du}{u}
\eeq

\subsection{ Massless final quark}

There are two possible ways of obtaining the $J(x; Q^2)$ distribution from the inverse Mellin trasform.
Each of them has its own peculiarities, since we are dealing with truncated expressions.
We have used both in order to compare the results and increase their reliability.

One way is to use an analytical expression for the inverse Mellin transform  (\ref{inv-m2}).
 Indeed, the massless form factor $J_N (Q^2)$  is
\begin{equation}
J(x; Q^2)=-x \,\frac{d}{dx} \Big\{\theta\left( 1-x-\epsilon \right)\, \Sigma
\left( x; \, Q^2 \right) \Big\} \qquad \qquad \textrm{at} \quad \epsilon \rightarrow 0,
\label{first-term}
\end{equation}
where  $ \Sigma(x; Q^2)$ is
the  inverse Mellin transform of $J_N/ N$.

The $\theta(1-x)$ function ensures the unitary normalization of the
distribution in the interval $(0,1)$ and it can be omitted in
the massless case since the function is regular at the origin.

We have, at NNLL \cite{Aglietti:2002ew} that :
\begin{eqnarray}
 \Sigma(x; Q^2) &=& \frac{e^{F_0(l)}}{\Gamma(1-F_1^{NL})}\left[ 1 + {F_1^{N^2
L}}\, \psi (1- {F_1^{NL}}) + \frac{1}{2}\, F_2(l)\, \left(\psi^2(1- {F_1^{NL}%
}) - \psi^\prime(1- {F_1^{NL}}) \right)\right]  \notag \\
&=& \frac{e^{l\, g_1(\beta_0 \alpha_s l) + g_2( \beta_0 \alpha_s l) +
\alpha_s \, g_3 (\beta_0 \alpha_s l)}}{ \Gamma( 1- g_1(\beta_0 \alpha_s l) -
\beta_0 \alpha_s \, l \, g_1^\prime(\beta_0 \alpha_s l))} \left[1+ \beta_0
\alpha_s \, g_2^\prime(\beta_0 \alpha_s l) \, \psi ( 1- g_1(\beta_0 \alpha_s
l) - \beta_0 \alpha_s \, l \, g_1^\prime(\beta_0 \alpha_s l)) \right.  \notag
\\
& & \left. +\frac{1}{2}\, F_2(l)\, \left(\psi^2 ( 1- g_1(\beta_0 \alpha_s l)
- \beta_0 \alpha_s \, l \, g_1^\prime(\beta_0 \alpha_s l))- \psi^\prime( 1-
g_1(\beta_0 \alpha_s l) - \beta_0 \alpha_s \, l \, g_1^\prime(\beta_0
\alpha_s l)) \right)\right]\notag \\
\label{Gfull}
\end{eqnarray}
where
\begin{eqnarray}
F_0(l) &=& l\, g_1(\beta_0 \alpha_s l) + g_2( \beta_0 \alpha_s l) + \alpha_s
\, g_3 (\beta_0 \alpha_s l),  \notag \\
F_1^{NL}(l) &\equiv& g_1(\beta_0 \alpha_s l) + \beta_0 \alpha_s \, l \,
g_1^\prime(\beta_0 \alpha_s l)  \notag \\
F_1^{N^2 L}(l) &\equiv& \beta_0 \alpha_s \, g_2^\prime(\beta_0 \alpha_s l).
\notag \\
F_2(l) &=& 2 \beta_0 \alpha_s \, g_1^\prime(\beta_0 \alpha_s l) + \beta_0^2
\alpha_s^2\, l \, g_1^{\prime\, \prime}(\beta_0 \alpha_s l).  \notag
\end{eqnarray}

Here $\Gamma$ is the Euler Gamma function, $\psi(x)= d\log\Gamma(x)/dx$, the digamma function, and $l\equiv-\ln(-\ln x)$. Note that $l
\rightarrow -\ln (1-x) $ when $x \rightarrow 1$.

Expression (\ref{Gfull}) can be rewritten in a synthetic way by evidencing the NLL part, that is as
\begin{equation}
\label{sigma} \Sigma \left(x; \, Q^2 \right) \,=\, \frac{ e^{ l\,
g_1(\tau) \, +  \, g_2(\tau)  }  } { \Gamma\left[1 - h_1(\tau)
\right]  } \,\,\delta\Sigma
\end{equation}
with
\beq \tau \equiv \beta_0 \alpha_s  l \, , \qquad  h_1(\tau)\equiv \frac{d}{d\tau} (\tau \, g_1(\tau)
) \eeq
and
\beq
\delta\Sigma = K_1 \, \,
e^{ \alpha_s \, g_3(\tau) }
\Bigg\{
1 \, + \, \beta_0 \, \alpha_s \, g_2'(\tau) \,
\psi\left[ 1 - h_1(\tau) \right]
\, + \, \frac{1}{2} \beta_0 \, \alpha_s \, h_1'(\tau)
\Big\{
\psi^2\left[1-h_1(\tau)\right]
- \psi'\left[1-h_1(\tau)\right]
\Big\}
\Bigg\} .
\eeq
Here, $K_1$ is a normalization factor such that $\delta\Sigma \to 1$ when $l \to 0$ (or $x \rightarrow 0$).

Another possibility is to obtain the inverse Mellin transform numerically,
by integrating Eq.~\ref{gen-resumming1}) at next-to-leading order.

This numerical integral is not straightforward, since, as we have seen, the $g_i$ are singular in $\lambda$ and their singularity reflects into $N$.
In other terms, the numerical distribution is not real for any value of N because of the integration over the Landau pole. An exact numerical evaluation of the inverse transform then requires a prescription for the pole. We use the  Minimal Prescription (MP), on a suitable path to the left of all the singularities \cite{Catani:1996}.
We have compared the analytical distribution (\ref{first-term}) with the distribution
obtained numerically.  The two curves show a very good agreement, although they differ slightly around $x \sim 1$, since the analytical ones reach the peak and start oscillating earlier-

\subsection{Mass correction factor}
\label{subsec4}

In analogy to the massless case, the mass-correction factor in physical space is obtained by means
of the derivative of the inverse Mellin transform of $\delta_N/N$: \beq \delta\left(
x;\, Q^2, m^2 \right) \, = \, -x \, \frac{d}{dx} \,
\left\{\int_{c-i\infty}^{c+i\infty} \frac{d N}{2\pi i N} \,
x^{-N} \, \delta_N\left( Q^2, m^2 \right) \right\}, \eeq where
$c$ is a (real) constant chosen in such a way that the integration
contour lies to the right of all the singularities of $\delta_N$.
By defining  \beq \bar{\delta}_{N r} \, \equiv \,\delta_{N} \eeq
and by changing variable from $N$ to $\nu = N r$, we obtain: \beq
\delta\left( x; \, Q^2, m^2 \right) \, = \,-x\, \frac{d}{dx} \,
\left\{ \int_{c \, r - i\infty}^{c \, r + i\infty}
\frac{d\nu}{2\pi i\nu} \, x^{-\nu/r} \,
\bar{\delta}_{\nu} \left( Q^2, m^2 \right) \right\}, \eeq

After this change of variable we can neglect the $\theta[\nu-1]$
and make the analytic continuation in the complex N-space.

We can therefore use the results in \cite{Aglietti:2002ew} to
obtain the correction factor in physical space in NNLL
approximation: \beq \delta\left( x; \, Q^2, m^2 \right) \, = \,
-x\,\frac{d}{dx} \, \Big\{ \theta( 1-x -\epsilon ) \, \Delta \left( x; \,
Q^2, m^2 \right) \Big\} \qquad \qquad \textrm{at} \quad \epsilon \rightarrow 0,
\label{second-term} \eeq where:
\begin{equation}
\label{SigmaNNLO} \Delta \left( x; \, Q^2, m^2 \right) \,=\,
\frac{  e^{ l^\prime \, d_1(\tau^\prime) \, +  \, d_2(\tau^\prime)  }  } {
\Gamma\left[1 - h_1(\tau^\prime) \right]  } \delta\Delta
\end{equation}
is the resummed partially integrated (or cumulative ) form factor
and the $\theta(1-x-\epsilon)$ ensures the unitary normalization
of the distribution in the interval $(0,1)$.  As already observed,
this term can be omitted in the massless case since the function
is regular at the boundary, but protects the mass
correction factor which is not a regular physical distribution.

In Eq.~(\ref{SigmaNNLO}) we have defined
\begin{equation}
l^\prime \equiv  -\log (-\log x^{1/r}) \qquad \qquad
\tau^\prime\equiv \beta_0 \alpha_s l^\prime .
\end{equation}
and
\begin{equation}
h_1(\tau^\prime) \, \equiv \, \frac{d}{d\tau^\prime}\left[\lambda d_1(\tau^\prime)\right]
\, = \, d_1(\tau^\prime) + \lambda \, d_1'(\tau^\prime).
\end{equation}
$\delta \Delta$ is a NNLL correction factor which can be set
equal to one in NLL:
\begin{equation}
\delta\Delta_{NLL} \, = \, 1.
\end{equation}
Its NNLL expression reads:
\begin{equation}
\delta\Delta \,=\, \frac{ S ~ }{ ~~ S|_{L\rightarrow 0} }
\end{equation}
with
\begin{equation}
S \,=\, e^{ \alpha_s \, d_3(\tau^\prime) }
\Bigg\{
1 \, + \, \beta_0 \, \alpha_s \, d_2'(\tau^\prime) \,
\psi\left[ 1 - h_1(\tau^\prime) \right]
\, + \, \frac{1}{2} \beta_0 \, \alpha_s \, h_1'(\tau^\prime)
\Big\{
\psi^2\left[1-h_1(\tau^\prime)\right]
- \psi'\left[1-h_1(\tau^\prime)\right]
\Big\}
\Bigg\}.
\end{equation}
$\Gamma(x)$ is the Euler Gamma function and
\begin{equation}
\psi(x) \, \equiv \, \frac{d}{d x} \log \Gamma(x)
\end{equation}
is the digamma function.

 It is
convenient to approximate the argument of the inverse Mellin
transform for $y\equiv 1-x \ll r$ by the expansion: \beq
\left[(1-y)^{1/r}\right] \, \simeq \,
 1-\frac{y}{r} \, + \, O\left( \frac{y^2}{r^2}\right), \eeq
 so that

 \beq \delta\left( y; \, Q^2, m^2 \right) \, = \, (1-y) \, \frac{d}{dy} \,
\left\{ \int_{c \, r - i\infty}^{c \, r + i\infty}
\frac{d\nu}{2\pi i\nu} \, \left[ ~ 1-\frac{y}{r} \right]^{-\nu}
\bar{\delta}_{\nu} \left( Q^2, m^2 \right) \right\}. \eeq

Note that the r.h.s. is positive only for $y<r$, implying that the
linearization above shrinks the domain of $y$ from the unitary
interval $(0,1)$ to the much smaller interval $(0,r)$. The
correction factor in physical space is therefore the inverse
Mellin transform of $\bar{\delta}_\nu$ with respect to
$(1-\frac{y}{r})$

 In this case it is useful to
employ the limit definition as for the plus-distribution defined in
Ref.\cite{Aglietti:2005mb}, such as:
 \beq  \delta\left( y; \, Q^2, m^2 \right) \,=  \lim_{\,\,\, \epsilon\rightarrow 0^+}  \,
(1-y)\,\frac{d}{dy} \, \Big\{\theta\left( y-\epsilon \right) \,
\Delta \left( y; \, Q^2, m^2 \right) \Big\} \eeq

 Finally: \beq \tau^\prime \, = \, \beta_0
\alpha_s \, L \eeq and \beq L \, = \, - \, \log
\left[-\log\left[1 - \frac{y}{r} \right]\right]. \eeq  A further
approximation step can be made in order to obtain the
final result:\beq -\log\left[1-\frac{y}{r}\right] \, \simeq \,
 \frac{y}{r}+ \, O\left( \frac{y^2}{r^2}\right) . \eeq

Finally the resummed expression in physical space reads: \beq
\delta\left( y; \, Q^2, m^2 \right) \,=  \lim_{\,\,\,
\epsilon\rightarrow 0^+}  \, (1-y)\,\frac{d}{dy} \,
\Big\{\theta\left( y-\epsilon \right) \, \Delta \left( y; \, Q^2,
m^2 \right) \Big\} \eeq  where $\Delta\left( y; \, Q^2, m^2
\right)$ is given by Eq.~(\ref{SigmaNNLO}) and \beq L \, = \,
\theta(r-y) \, \log \frac{r}{y}.\eeq We have limited the
domain to $y < r$ with a $\theta$-function\footnote{ As suggested
in Ref. \cite{nostro2007-1} a smooth approximation to the $ Theta
\otimes Log $ function form can be given by: $L \, \simeq \, - \, \log \left[ 1 - (1 -
y)^{1/r} \right]$. In fact these functions agree at the first
order approximation.}.


We are now ready to perform the convolution in order to obtain the physical distribution.
The physical form distribution is obtained by the Mellin transform of
Eq.~(\ref{univ}), that is by
 \beq J\left(
x;\, Q^2, r \right) \, = \,
\int_{c-i\infty}^{c+i\infty} \frac{d N}{2\pi i} \,
x^{-N} \,J_N(Q^2; \,r) = \, \int_{c-i\infty}^{c+i\infty} \frac{d N}{2\pi i} \,
x^{-N} J_N(Q^2) ~ \delta_N(Q^2; \,r)  \label{numeric1} \eeq

This integral is not straightforward, since, as we have seen, the $g_i$ are singular in $\lambda$ and their singularity reflects in $N$.

$J\left(
y;\, Q^2, r \right) $ can also be computed analytically, by the convolution of the inverse Mellin transforms of
$J_N(Q^2) $ and $ \delta_N(Q^2; \,r)$.

\beq J\left(x;\, Q^2, r \right) \, = \, \int_x^{1} \frac{d
z}{z} \, J(z; Q^2) \, \delta\left(\frac{x}{z};Q^2,r \right) \,
\eeq where $J\left(y; \, Q^2 \right)$ and $\delta\left(y; \, Q^2, r \right)$
are given by Eq.~(\ref{first-term}) and Eq.~(\ref{second-term}), respectively.
Therefore, we obtain the following analytical expression:
\beq  J\left(x;\, Q^2, r \right) \, = \, \int_{x}^{1} \frac{d
z}{z} \, J(z; Q^2) \, \delta\left(\frac{x}{z};Q^2,r \right)
\, \eeq
 \beq = \, \, \lim_{\epsilon\rightarrow 0^+} \, \int_{x}^{1} dz
\frac{x}{z} \, \left\{-\delta(1-z-\epsilon)\,
\Sigma^{'}(\frac{x}{z}; Q^2) \,\, \Delta\left(z;Q^2,r \right) \,+
\theta(1-z-\epsilon) \Sigma^{'}(\frac{x}{z}; Q^2) \,\,
\Delta^{'}\left(z;Q^2,r \right)\right\} \nonumber
\label{convoluzione}\eeq
where we have the Dirac delta function  $\delta(1-z-\epsilon) = -d \theta(1-z-\epsilon)/dz $.
Let us observe that $ \Sigma (y; Q^2) \rightarrow 0$ when $y \rightarrow 0$.

\section{Frozen Coupling approximation}
\label{FC-section}

The frozen coupling approximation means neglecting the variation of $\alpha_s
$ with the scale. We first look for solution in the frozen coupling approximation;
in the massless case  the resumming  formula at NNLL gives:
\begin{eqnarray}
\log J_{N}\left(Q^2\right) &=& \int_{0}^{1}dx\,\frac{x^{N-1}-1}{1-x}
\Big\{ \int_{Q^{2} (1-x)^{2}}^{Q^{2} (1-x)}\frac{dk^{2}}{k^{2}} \,\left[ A_1
\alpha_s+ A_2 \alpha_s^2 + A_3 \alpha_s^3 + ... \right] +
\notag \\
&+& B_1 \alpha_s + B_2 \alpha_s^2 + ... + D_1 \alpha_s + D_2
\alpha_s^2 + ... + \Big\}  \notag \\
&\simeq& \int_{0}^{1} dx \frac{x^{N-1}-1}{1-x} \Big\{ \left( A_1
\alpha_s+ A_2 \alpha_s^2 + A_3 \alpha_s^3 \right) \ln \frac{1}{1-x} +   \notag \\
&+& \left( B_1 + D_1 \right) \alpha_s + \left( B_2 + D_2 \right) \alpha_s^2
\Big\}  \notag \\
& &
\end{eqnarray}

  In the frozen coupling approximation,  $\beta_0
\rightarrow 0$. After integration in $z$, we expand in $\lambda=
\beta_0 \alpha_s \, L$ where  $L=\log N $. At the lowest order, in
the massless case, we have:
\begin{eqnarray}
g_1 &=& - \frac{A_1}{2 \, \beta_0} \, \lambda \\
g_2 &=& \left( -\frac{B_1}{\beta_0} -\frac{D_1}{\beta_0} -\frac{A_1 \gamma_E%
}{\beta_0} \right) \, \lambda \\
g_3 &=& \left( -\frac{B_2}{\beta_0} -\frac{D_2}{\beta_0} -\frac{A_2 \gamma_E%
}{\beta_0} \right) \, \lambda
\end{eqnarray}

We can easily find the corresponding $d_i$ for the mass correction term  by the substitution
$A_i  \rightarrow -A_i$,  $B_i  \rightarrow -B_i$ and $\lambda \rightarrow \rho$.

Disregarding the NLL terms we have in the
frozen coupling limit:
\begin{eqnarray}
\log J_N(Q^2,m^2) &=&  f_N(Q^2)+ F_N(Q^2,m^2)  \nonumber \\
 & \simeq & L g_1 + L_r d_1
 \nonumber \\
 &=& - \frac{A_1}{2 \, \beta_0} \,  (\lambda \, L- \rho \, L_r)
\label{logapp}
\end{eqnarray}
where $\rho=\alpha_s \beta_0
\, L_r$ with $L_r=\log N r$. Then the
leading behavior for the mass corrected formula is determined by:
 \be \log J_N(Q^2,m^2) \simeq A_1 \,  \alpha_s \,
 \log N \, \log r
 \label{logdom}
\ee
The divergent double logarithmic behavior for $N \rightarrow \infty$  of the massless case is replaced by a single logarithm times
a new regularizing term $ \log r$.
This term is negative and restores a finite  limit when $ x \rightarrow 1 $  in the physical space.
The peculiarity of  applying the MP is that due to the milder singularity,
we no longer have  the factorially growing spurious contributions described in \cite{Catani:1996}
 generated by neglecting certain subleading terms when the moment space formula is turned to an $x$-space formula.
The resummed massive case distribution is then a regular function
in the limit $x\rightarrow 1$. We have therefore found  that, also in the massive case, the resummed formula is
 void of unwanted spurious ambiguities.

\begin{figure}[htb]
\includegraphics[width=21pc]{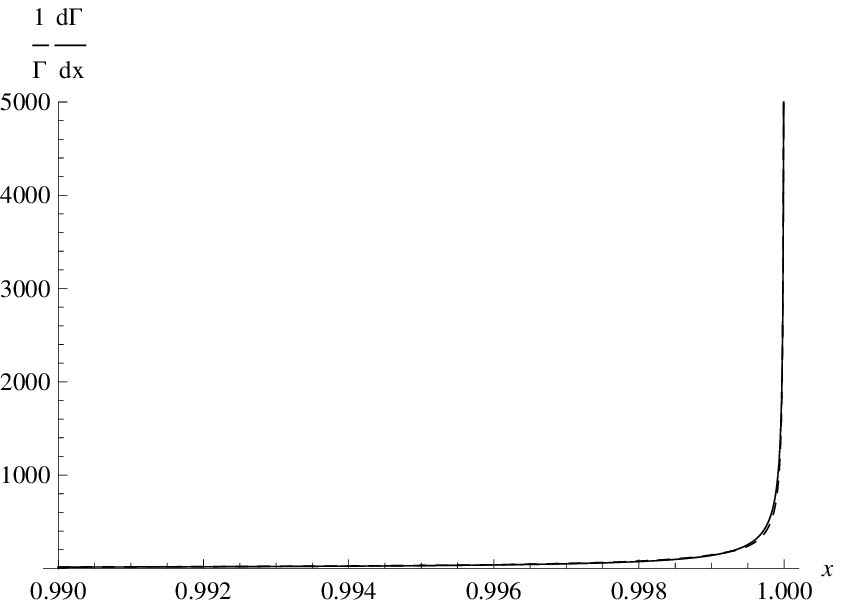}
\includegraphics[width=21pc]{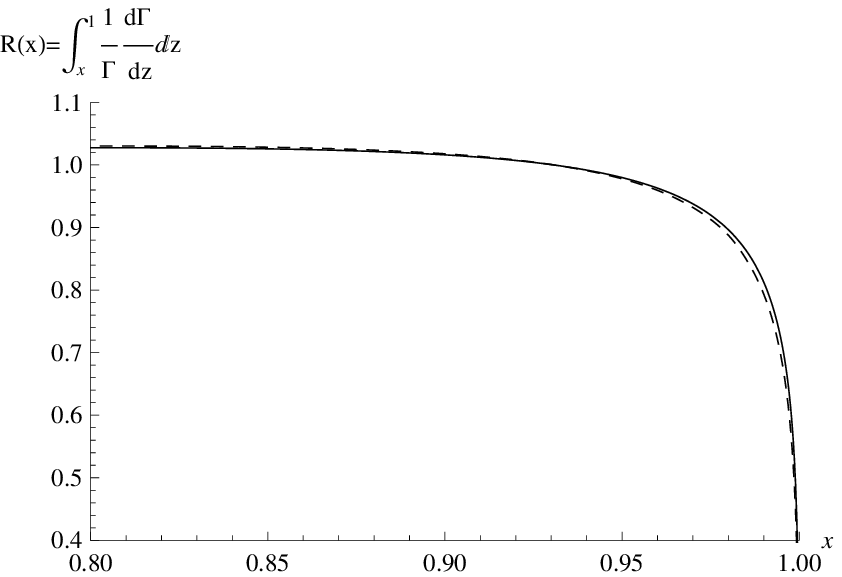}
\caption{Frozen coupling approx.: top decay jet rates (on the left) and partially
integrated jet rates (on the right). Comparison between massless
(continuous line) and massive (dashed line) distributions.}
\label{toptobjr}
\end{figure}

\begin{figure}[htb]
\includegraphics[width=21pc]{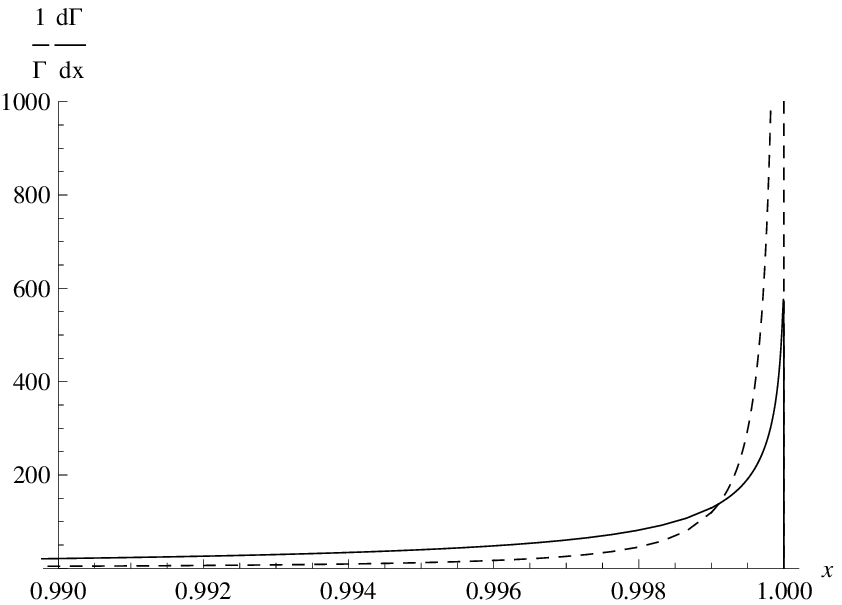}
\includegraphics[width=21pc]{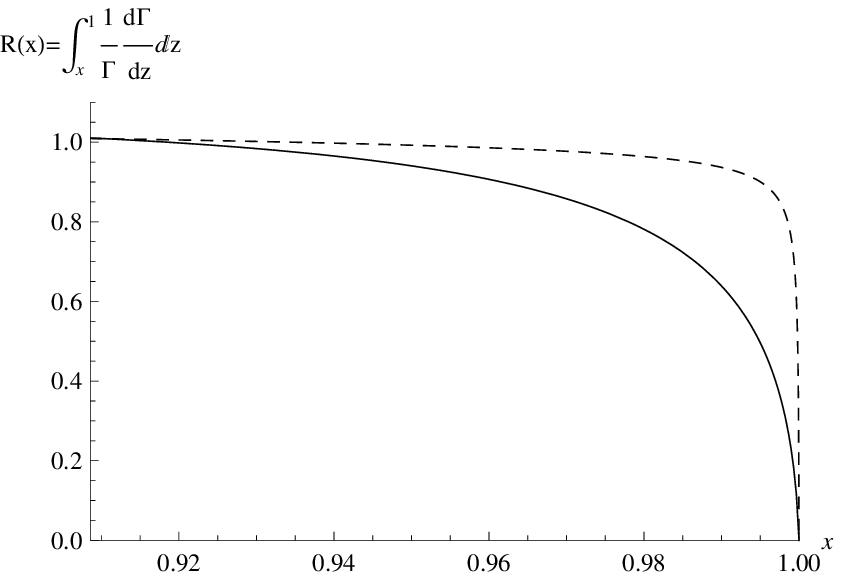}
\caption{Frozen coupling approx.: bottom decay jet rates (on the left) and partially integrated
jet rates (on the right). Comparison between massless (continuous line)
and massive (dashed line) distributions.} \label{btocjr}
\end{figure}

It has been already mentioned that there are  two ways, analytical and
numerical, to compute the inverse Mellin transforms of $J_N(Q^2;
r)$. By numerical method we mean the direct numerical integration
in Eq.~(\ref{numeric1}); by analytical method we mean to use the
approximated analytical expression for the convolution in
Eq.~(\ref{convoluzione}). In the frozen coupling case, the $g_i$
and $d_i$ are linear in $\lambda$ and therefore the numerical path
does not include the Landau pole; the numerical integration
becomes therefore exact.

In figs. \ref{toptobjr}-\ref{btocjr} we compare the (normalized)
resummed massless and massive jet rates, in the frozen
coupling approximation,  for  top to b  and  for  b to c  decays, respectively. We obtain the same results by calculating the jet factor with both numerical and analytical methods.

\section{Resumming with a running coupling constant}
\label{conclusionsec}

\begin{figure}[htb]
\begin{center}
\includegraphics[width=21pc]{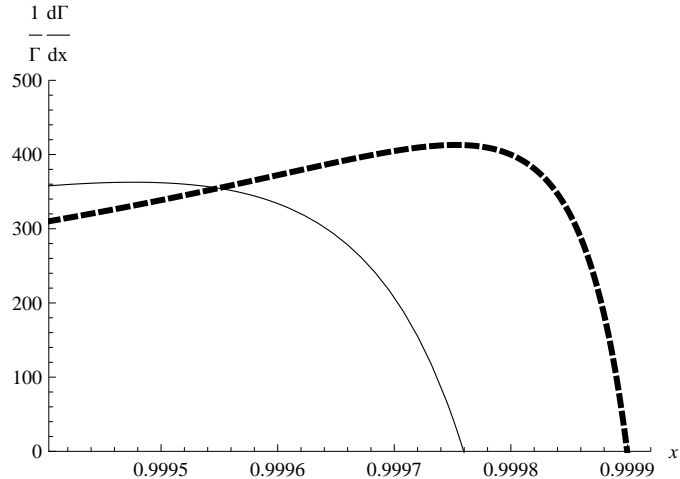}
\caption{Top decay jet rates: the massive case (dashed line) compared to
the massless case (continuous line) at NNLL; $\alpha_s$ is set to
the top mass scale, $\alpha_s=0.11$.}
\label{top-running}
\end{center}
\end{figure}

\begin{figure}[htb]
\begin{center}
\includegraphics[width=21pc]{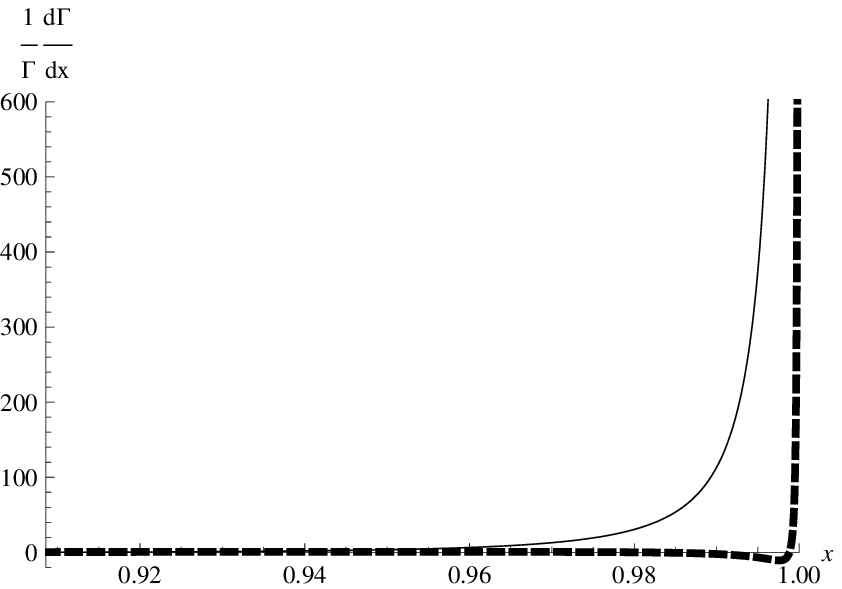}
\caption{B decay jet rates: the massive case (dashed line) compared to
the massless case (continuous line) at NNLL; $\alpha_s$ is set to
the beauty mass scale, $\alpha_s=0.219$.}
\label{bottom-running}
\end{center}
\end{figure}

\begin{figure}[htb]
\begin{center}
\includegraphics[width=21pc]{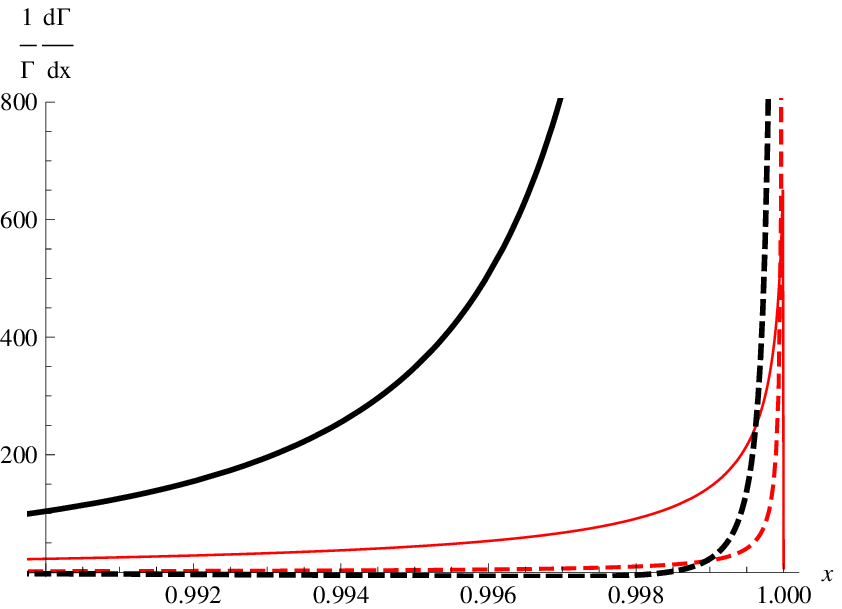}
\caption{Comparison between the frozen coupling  (lighter/red lines) and the running coupling case
 (thicker/black lines) in B decay jet rates. The dashed (continuous) lines refers the massive (massless) case, at NNLL and with $\alpha_s=0.23$.}
\label{comparisonFR}
\end{center}
\end{figure}

By releasing the frozen coupling approximation, we can calculate the distribution with a running coupling.
We need in this case a regularization procedure. Indeed, in the massive case,
 we have four poles on the real axis of the Mellin complex
plane respectively for $ \lambda=\frac{1}{2}$, $\lambda=1 $ and for $
\rho=\frac{1}{2}$, $\rho=1 $. These poles arise from the two
logarithmic structures of the massless and massive correction
formulas. The important point is that the massive poles stand to the right of the
massless ones, therefore  allowing to use the MP procedure of Ref. \cite{Catani:1996}.
We have integrated numerically Eq.~(\ref{numeric1}), over the
path  made by two straight lines parallel to the negative real axis, closed by a half-circle centered around
the origin and crossing the positive axis between the origin and the first Landau pole.
The integrated expression is computed at NNLL order, releasing the frozen coupling approximation;  the coupling runs over the whole integration range.

In fig. \ref{top-running} we compare the resummed massless and massive jet rates for the case of  $t$  to $b$ decays. All the plots in this section are not normalized. The massless plots coincide, in the considered range, with the ones obtained by the analytical distribution of Eq.~(\ref{first-term}).
 Differences between massless and massive case start, as expected, approximately for values of $x \simeq 1-r \simeq 0.999$\footnote{We have, approximately, $z^{N-1}-1 \sim -\theta(1-z-1/N)$ \cite{catanitrentadue}, and we can set $x \leq 1-1/N \simeq 1-r .$}. We have only listed the NNLL  plots, since they do not differ, substantially, with respect to NLL ones. The MP does not avoid approaching to the essentially non-perturbative regime  at $x \simeq 1$, where the plots start to oscillate and, therefore,  a physically motivated treatment of non-perturbative effects has to be introduced.
We find that this physical non perturbative cut-off can be put, in practice,  equal to 1 in top decay.

In the b decay case (fig. \ref{bottom-running}),  the two curves start differentiating at $x \simeq 1-r \simeq 0.9$, and the difference is visible, with the massive case being less divergent, as expected. The rates in the curves are not normalized, and the NNLL plot  presents relevant difference from the NLL one; the addition of NNLL terms renders the plot closer in the rising behavior to the massless plot. At NNLL order we deal not only with double logarithms, but also with single logarithms that shift the position of the minimum \footnote{see f.i. formula (62) in  \cite{Aglietti:2005mb}.}.
The effect of the running coupling, affecting in a very distinct manner the massless and the massive case,  can  be observed by comparing with fig. \ref{btocjr}.
In order to facilitate the comparison with the frozen coupling case, we report both the massless and massive distributions, not normalized, with frozen and running couplings, in fig. \ref{comparisonFR}.

\section{Conclusions}
\label{conclusionfin}

In this paper we have considered the application of the MP to the resummed jet function distributions for heavy quark decays
in massless and massive final quarks.
We have used a QCD resummation formula in $\alpha_s$, that takes into account contribution from large collinear and soft logarithms near the threshold region \cite{nostro2007-1}.
Benefits of large threshold logarithms resummation are restoring predictive power to the perturbation series and
increase theoretical accuracy, f.i. by a  reduction of scale
uncertainty. Accuracy is obviously increased by including higher order terms in the
exponent. We have considered NNLL order corrections. In the case of massive final quarks, we have analyzed the applicability of  the Minimal Prescription scheme to deal with the unavoidable problem, connected with any physical application of QCD resumming formulas, of the integration over the Landau pole.

We calculated the form factors in Mellin and physical space, in order to
extract the main and universal features of the perturbative distributions, common to all processes at the threshold.
We consider this a necessary first step to approach the phenomenological study of specific decay rates in the QCD resummed framework. In order to compare with data,  the further step is to
match the distribution with a fixed order coefficient function and include non perturbative effects \cite{future}

\noindent
As a final remark, let us notice that
  a very interesting application is to the inclusive semi-leptonic $b$ decay into $c$. In such partonic three body decay, one can combine the resummed with the full  triple differential distribution.

\noindent
We have considered top quark decays, for the recently increased interest due to the large amount of data available at the hadronic machines. One possible application of the present results is to explore the effects of the resummation on the calculation  of helicity fractions of the W boson from top quark decays. These are presently measured with increasing precision by both the CDF and the D0 collaboration at the Tevatron.

\section{Acknowledgements}
We would like  to thank U. Aglietti and G. Corcella for very useful discussions.
 One of us (L.D.G.)   thanks the
Fondazione A. Della Riccia (Firenze) for financial support and the Theory Group at SLAC for their warm hospitality.

\section{Appendix}
\label{appendix}

In this appendix we enlist  notations and actual values used in the paper.

The functions $g_{1}$ and $g_{2}$  introduced in section (\ref{subsec3}) have the following expressions \cite
{americani}:
\begin{eqnarray}
g_{1}\left( \lambda ;\frac{\mu ^{2}}{Q^{2}}\right) &=&-\frac{A_{1}}{2\beta
_{0}}\,\frac{1}{\lambda }\left[ \left( 1-2\lambda \right) \log \left(
1-2\lambda \right) -2\left( 1-\lambda \right) \log \left( 1-\lambda \right) %
\right] ; \nonumber  \\
\,g_{2}\left( \lambda ;\frac{\mu ^{2}}{Q^{2}}\right) &=&+\frac{A_{2}}{2\beta
_{0}^{2}}\left[ \log (1-2\lambda )-2\log (1-\lambda )\right] +\frac{%
A_{1}\gamma _{E}}{\beta _{0}}\left[ \log (1-2\lambda )-\log (1-\lambda )%
\right] +  \notag \\
&&-\frac{\beta _{1}A_{1}}{4\beta _{0}^{3}}\left[ \log ^{2}(1-2\lambda
)-2\log ^{2}(1-\lambda )+2\log (1-2\lambda )-4\log (1-\lambda )\right] +
\notag \\
&&+\frac{D_{1}}{2\beta _{0}}\log (1-2\lambda )+\frac{B_{1}}{\beta _{0}}\log
(1-\lambda )+\frac{A_{1}}{2\beta _{0}}\,\left[ \log \left( 1-2\lambda
\right) -2\log \left( 1-\lambda \right) \right] \log \frac{\mu ^{2}}{Q^{2}} \nonumber
\end{eqnarray}

The NNLO function $\,g_{3}$\ \cite{Aglietti:2002ew} reads:
\begin{eqnarray}
g_{3}\left( \lambda ;\frac{\mu ^{2}}{Q^{2}}\right) &=&-\frac{A_{3}}{2\beta
_{0}^{2}}\left[ \frac{\lambda }{1-2\lambda }-\frac{\lambda }{1-\lambda }%
\right] -\frac{A_{1}\zeta _{2}}{2}\left[ \frac{4\lambda }{1-2\lambda }-\frac{%
\lambda }{1-\lambda }\right] +  \notag \\
&&-\frac{A_{1}\beta _{2}}{4\beta _{0}^{3}}\left[ \frac{2\lambda
}{1-2\lambda }-\frac{2\lambda }{1-\lambda }+2\log \left(
1-2\lambda \right) -4\log \left(
1-\lambda \right) \right] +  \notag \\
&&+\frac{A_{2}\beta _{1}}{2\beta _{0}^{3}}\left[ \frac{\log \left(
1-2\lambda \right) }{1-2\lambda }-\frac{2\log \left( 1-\lambda \right) }{%
1-\lambda }+\frac{3\lambda }{1-2\lambda }-\frac{3\lambda }{1-\lambda }\right]
+  \notag \\
&&-\frac{A_{1}\beta _{1}^{2}}{2\beta _{0}^{4}}\left[ \frac{1}{2}\frac{\log
^{2}\left( 1-2\lambda \right) }{1-2\lambda }-\frac{\log ^{2}\left( 1-\lambda
\right) }{1-\lambda }+\frac{\log \left( 1-2\lambda \right) }{1-2\lambda }%
+\right.  \notag \\
&&\left. -\frac{2\log \left( 1-\lambda \right) }{1-\lambda }+\frac{\lambda }{%
1-2\lambda }-\frac{\lambda }{1-\lambda }-\log \left( 1-2\lambda \right)
+2\log \left( 1-\lambda \right) \right] +  \notag \\
&&+\frac{D_{1}\beta _{1}}{2\beta _{0}^{2}}\left[ \frac{\log \left(
1-2\lambda \right) }{1-2\lambda }+\frac{2 \lambda }{1-2\lambda }\right] +%
\frac{B_{1}\beta _{1}}{\beta _{0}^{2}}\left[ \frac{\log \left( 1-\lambda
\right) }{1-\lambda }+\frac{\lambda }{1-\lambda }\right]+  \notag \\
&&-\frac{D_{2}}{\beta _{0}}\frac{\lambda }{1-2\lambda }-\frac{B_{2}}{\beta
_{0}}\frac{\lambda }{1-\lambda }-\frac{A_{1}\gamma _{E}^{2}}{2} \left[ \frac{%
4 \lambda }{1-2\lambda }- \frac{\lambda}{1-\lambda }\right] +  \notag \\
&&+\frac{A_{1}\beta _{1}\gamma _{E}}{\beta _{0}^{2}}\left[ \frac{\log \left(
1-2\lambda \right) }{1-2\lambda }-\frac{\log \left( 1-\lambda \right) }{%
1-\lambda }+\frac{1}{1-2\lambda }-\frac{1}{1-\lambda }\right] +  \notag \\
&&-\frac{A_{2}\gamma _{E}}{\beta _{0}}\left[ \frac{1}{1-2\lambda }-\frac{1}{%
1-\lambda }\right] -\frac{D_{1}\gamma _{E} 2 \lambda}{1-2\lambda }-\frac{%
B_{1}\gamma _{E} \lambda}{1-\lambda }+  \notag \\
&&-\frac{A_{1}}{2\beta _{0}}\left[ \frac{2\lambda ^{2}}{1-2\lambda }-\frac{%
\lambda ^{2}}{1-\lambda }\right] \log ^{2}\frac{\mu ^{2}}{Q^{2}}-\frac{A_{2}%
}{\beta _{0}^2}\left[ \frac{\lambda }{1-2\lambda }-\frac{\lambda }{1-\lambda
}\right] \log \frac{\mu ^{2}}{Q^{2}}+  \notag \\
&&-\frac{A_{1}\gamma _{E}}{\beta _{0}}\left[ \frac{2\lambda }{1-2\lambda }-%
\frac{\lambda }{1-\lambda }\right] \log \frac{\mu ^{2}}{Q^{2}}-\frac{D_{1}}{%
\beta _{0}}\frac{\lambda }{1-2\lambda }\log \frac{\mu ^{2}}{Q^{2}}-\frac{%
B_{1}}{\beta _{0}}\frac{\lambda }{1-\lambda }\log \frac{\mu ^{2}}{Q^{2}}+
\notag \\
&&+\frac{A_{1}\beta _{1}}{\beta _{0}^{3}}\left[ \frac{\lambda \log \left(
1-2\lambda \right) }{1-2\lambda }-\frac{\lambda \log \left( 1-\lambda
\right) }{1-\lambda }+\frac{\lambda }{1-2\lambda } + \right.  \notag \\
&& \left. -\frac{\lambda }{1-\lambda } + \frac{1}{2} \log (1-2 \lambda) -
\log (1-\lambda)\right] \log \frac{\mu ^{2}}{Q^{2}}.  \nonumber
\end{eqnarray}
Arbitrary constants have been added to the function $g_3$ in order
to make it homogenous. The quantity $\gamma _{E}=0.577216\ldots $
is the Euler constant and $\zeta \left( n\right) $ is the Riemann
zeta function,
\begin{equation}
\zeta \left( n\right) \equiv \sum_{k=1}^{\infty }\frac{1}{k^{n}}.\nonumber
\end{equation}
$\zeta \left( 2\right) =\pi ^{2}/6=1.64493.$ The functions $g_{2}$ and $%
g_{3} $ depend on the renormalization scale $\mu ,$ while $g_{1}$ does not.

The known values for the resummation constants defined in section (\ref{subsec3}) read:
\begin{eqnarray}
A_1 &=& \frac{C_F}{\pi};\nonumber
\\
A_2 &=&  \frac{C_F}{\pi^2} \left[ C_A\left( \frac{67}{36} - \frac{z(2)}{2} \right)
- \frac{5}{18} n_f \right]; \nonumber
\\
A_3 &=& \frac{C_F}{\pi^3}
\Bigg[
C_A^2\Big(
\,\frac{245}{96} + \frac{11}{24} z(3) - \frac{67}{36} \, z(2) + \frac{11}{8} \, z(4)
\Big)
\,+\nonumber\\
& -& \, C_A \, n_f \Big( \frac{209}{432} + \frac{7}{12} z(3) - \frac{5}{18}\,z(2) \Big) +
\nonumber\\
&-& \, C_F \, n_f \Big( \frac{55}{96} - \frac{z(3)}{2}  \Big)
- \frac{n_f^2}{108}
\Bigg]; \nonumber
\\
B_1 &=&  - \frac{3}{4} \frac{C_F}{\pi}; \nonumber
\\
B_2 & = & \frac{C_F}{\pi^2}
\Bigg[
C_A \left( - \frac{3155}{864} + \frac{11}{12}\,z(2) + \frac{5}{2} \, z(3) \right)\,+\nonumber\\
&-& C_F \left( \frac{3}{32} + \frac{3}{2} z(3) - \frac{3}{4} \, z(2) \right)
\,+\nonumber\\
&+& n_f \left( \frac{247}{432} - \frac{z(2)}{6} \right)
\Bigg]; \nonumber
\\
D_1 &=&  - \frac{C_F}{\pi}; \nonumber
\\
D_2 &=& \frac{C_F}{\pi^2} \left[
C_A \left( \frac{55}{108} - \frac{9}{4} z(3) + \frac{z(2)}{2} \right)
+ \frac{n_f}{54}\right], \nonumber
\end{eqnarray}
where $C_A=N_c=3$ is the Casimir of the adjoint representation.

The knowledge of the
quantities $A_1$, $A_{2}$, $B_1$ and $D_1$ is needed for
resummation at next-to-leading order.

The coefficients $A_1$, $B_1$ and $D_1$ are renormaliza\-tion-scheme independent, as they can be
obtained from tree-level amplitudes with one-gluon emission.
The higher-order coefficients are instead renormalization-scheme dependent
and are given in the $\overline{MS}$ scheme for the coupling constant
\footnote{A discussion about the scheme dependence of the higher
order coefficients $A_2,\,B_2,$ etc. on the coupling constant
can be found in \cite{Catani:1990rr}.}.

The coefficients $d_i$, defined in section (\ref{subsecMass1}), are :
\begin{eqnarray}
d_1(\rho) &=&  \frac{{A_1}}{2\,\beta_0\, \rho}
\Big[
\left( 1 - 2\,\rho \right) \,
       \log (1 - 2\,\rho)
     - 2\,  \left( 1 - \rho \right)
           \,\log (1 - \rho)
\Big]; \nonumber
\\
d_2(\rho)&=& \frac{{D_1}}{2\,{{\beta}_0}} \log (1 - 2\,\rho)
- \frac{B_1}
   {{\beta}_0} \,\log (1 - \rho )
- \frac{{A_2}}{ 2\,
     {{\beta}_0}^2} \,
\Big[
\log (1 - 2\,\rho ) -
       2\,\log (1 - \rho )
\Big]  + \nonumber
\\ \nonumber
&+&\frac{{A_1}\, {{\beta}_1} }{4\,{{{\beta}_0}}^3}\,
\Big[
       2\,\log (1 - 2\,\rho ) +
       {\log^2 (1 - 2\,\rho )} -
       4\,\log (1 - \rho ) -
       2\,{\log^2 (1 - \rho )}
\Big] \, \nonumber
      +
\\ \nonumber
&-& \frac{{A_1}\,{{\gamma }_E}  }
     {{{\beta}_0}}\,
\Big[
\log (1 - 2\,\rho ) -
       \log (1 - \rho )
\Big]
- \frac{A_1}{2 \, \beta_0}
\Big[
\log(1 - 2\,\rho )-2 \,\log (1 -
\rho )
\Big]
\, \log \frac{\mu^2}{m^2}. \nonumber
\end{eqnarray}
For the NNLO function $d_3$ we obtain:
\begin{eqnarray}
d_3(\rho) &=&
- \, \frac{D_2}{\beta_0} \,
\frac{\rho}{1 - 2\,\rho }
\, - \, D_1 \, \gamma_E \, \frac{2 \, \rho }{1 - 2\,\rho }
+ \frac{{D_1}\,{\beta_1}}{2\,
     {{\beta_0}}^2}\,
\left[
\frac{2\,\rho }{1 - 2\,\rho } +
       \frac{\log (1 - 2\,\rho )}
        {1 - 2\,\rho }
\right]
\, + \nonumber\\
&+& \frac{B_2}{\beta_0}
\, \frac{\rho }{ 1 - \rho }
\, + \, B_1 \, \gamma_E \, \frac{\rho }{1 - \rho }
\, + \nonumber\\
&-&
  \frac{ B_1 \, \beta_1 }{ {\beta_0}^2 } \,
\left[
\frac{\rho }{1 - \rho } +
       \frac{\log (1 - \rho )}{1 - \rho }
\right]
+ \frac{A_3}
     {2\,{{\beta_0}}^2}\,
\left[
\frac{\rho }{1 - 2\,\rho } -
       \frac{\rho }{1 - \rho }
\right]
+\nonumber\\
&+&  \frac{A_2\,
       {\gamma_E} }{
       {\beta_0}}\,
\left[
    \frac{2 \, \rho }{1 - 2\,\rho } -
         \frac{\rho}{1 - \rho }
\right]
+\nonumber\\
&-&
  \frac{ A_2 \, \beta_1 }{ 2\,{\beta_0}^3 }\,
\left[
\frac{3\,\rho }{1 - 2\,\rho } -
       \frac{3\,\rho }{1 - \rho } +
       \frac{\log (1 - 2\,\rho )}
        {1 - 2\,\rho } -
       \frac{2\,\log (1 - \rho )}{1 - \rho }
\right]
+\nonumber\\
&+&
  \frac{ A_1 \, {\gamma_E}^2 }{2}\,
\left[
\frac{4\,\rho }{1 - 2\,\rho } -
       \frac{\rho }{1 - \rho }
\right]
+ \frac{ A_1 \,{\pi}^2 }{12}\,
\left[
\frac{4\,\rho }{1 - 2\,\rho } -
       \frac{\rho }{1 - \rho }
\right]
+\nonumber\\
 &+&
  \frac{ A_1 \, \beta_2 }{4\,{\beta_0}^3}\,
\left[
\frac{2\,\rho }{1 - 2\,\rho } -
       \frac{2\,\rho }{1 - \rho } +
       2\,\log (1 - 2\,\rho ) -
       4\,\log (1 - \rho )
\right]
+\nonumber\\
&-& \frac{ A_1 \, \beta_1 \, \gamma_E }{ {\beta_0}^2 }\,
\left[
\frac{2\, \rho}{1 - 2\,\rho } -
       \frac{\rho}{1 - \rho } +
       \frac{\log (1 - 2\,\rho )}
        {1 - 2\,\rho } -
       \frac{\log (1 - \rho )}{1 - \rho }
\right]
+\nonumber\\
&+&
  \frac{A_1\,{\beta_1}^2}{ 2\, {\beta_0}^4 }\,
\left[
\frac{\rho }{1 - 2\,\rho } -
       \frac{\rho }{1 - \rho } -
       \log (1 - 2\,\rho ) +
       \frac{\log (1 - 2\,\rho )}
        {1 - 2\,\rho }
\right. \nonumber \\
&+&
\left.
       \frac{ \log^2 (1 - 2\,\rho ) }
        {2\,\left( 1 - 2\,\rho  \right) } +
       2\,\log (1 - \rho ) -
       \frac{2\,\log (1 - \rho )}
        {1 - \rho } -
       \frac{ \log^2 (1 - \rho ) }{1 - \rho }
\right]
\, + \, \nonumber\\
&-&
  \frac{D_1}{\beta_0}
\frac{\rho }{ 1 - 2\,\rho } \,
     \log \frac{{\mu }^2}{m^2}
+ \frac{ B_1}{\beta_0}
\frac{\rho }{ 1 - \rho }\,
     \log \frac{{\mu }^2}{m^2}
+ \frac{{A_2}}{{{\beta_0}}^2}\,
\left[ \frac{\rho }{1 - 2\,\rho } -
       \frac{\rho }{1 - \rho }
\right] \,
     \log \frac{{\mu }^2}{m^2}
+\nonumber\\
&+& \frac{{A_1}\,
     {\gamma_E} }{{\beta_0}}\,
\left[
\frac{2\,\rho }{1 - 2\,\rho } -
       \frac{\rho }{1 - \rho }
\right] \,
     \log \frac{{\mu }^2}{m^2}
+\nonumber\\
&-& \frac{{A_1}
     {\beta_1} }{{{\beta_0}}^3}
\left[
\frac{\rho }{1 - 2\,\rho } -
       \frac{\rho }{1 - \rho } +
       \frac{\log (1 - 2\,\rho )}{2} +
       \frac{{\rho }\,
          \log (1 - 2\,\rho )}{1 - 2\,\rho }
\right.
+ \, \nonumber\\
&-&
\left.
\log (1 - \rho )
- \frac{\rho \,\log (1 - \rho )}
        {1 - \rho }
\right]
     \log \frac{{\mu }^2}{m^2}\!
+\nonumber\\
&+& \frac{{A_1}}{2\,
     {\beta_0}}\,
\left[
\frac{2\,{\rho }^2}
        {1 - 2\,\rho } -
       \frac{{\rho }^2}{1 - \rho }
\right] \,
     {\log^2 \frac{{\mu }^2}{m^2}}. \nonumber
\end{eqnarray}
The coefficients $\beta_i$ of the QCD $\beta$-function in our normalization
have been given in \cite{Aglietti:2005mb}.


\begin{thebibliography}{99}

\bibitem{parpet} G. Parisi and R. Petronzio, Nucl. Phys.   B 154, 427 (1979);
                 G. Curci and M. Greco, Phys. Lett. B92, 175 (1980).

\bibitem{kodtren} J. Kodaira and L. Trentadue, SLAC-PUB-2934 (1982);
                  Phys. Lett. B 112, 66 (1982);
S.~Catani, E. D'Emilio and L. Trentadue, Phys. Lett. B 211, 335 (1988).

\bibitem{cattren}  S. Catani and L. Trentadue, Nucl. Phys. B 327, 323 (1989).

\bibitem{cattren2} S. Catani and L. Trentadue, Nucl. Phys. B 353, 183 (1991).


\bibitem{sterman} G. Sterman, Nucl. Phys. B 281, 310 (1987).


 \bibitem{aglietti:2001}
 U.~Aglietti,
  Nucl.\ Phys.\ B  610 (2001)  293

\bibitem{Aglietti:2005mb}
U.~Aglietti, G.~Ricciardi and G.~Ferrera,
Phys.\ Rev.\ D  74 (2006) 034004

\bibitem{nostro2007-1}
U. Aglietti, L. Di Giustino,  G. Ferrera and L. Trentadue,
Phys. Lett. B 651 (2007) 275.

\bibitem{Catani:1996}
S. Catani, M. Mangano, P. Nason and L. Trentadue, Nucl. Phys. B 478 (1996) 273.

\bibitem{Gambino:2011fz}
  P.~Gambino, C.~Schwanda,
  [arXiv:1102.0210 [hep-ex]].

\bibitem{Cacciari:2002re}
M.~Cacciari, G.~Corcella and A.~D.~Mitov,
JHEP 0212 (2002) 015.



\bibitem{catcac}
M.~Cacciari and S.~Catani,
Nucl.\ Phys.\ B  617 (2001) 253.

\bibitem{catanitrentadue}
S.~Catani and L.~Trentadue,
Nucl.\ Phys.\ B  327 (1989) 323.

\bibitem{Catani:1990rr}
S.~Catani, B.~R.~Webber and G.~Marchesini,
Nucl.\ Phys.\ B  349 (1991) 635.



\bibitem{sghedoni}
U.~Aglietti, R.~Sghedoni and L.~Trentadue,
Phys.\ Lett.\ B  522 (2001) 83;
Phys.\ Lett.\ B  585 (2004) 131.

\bibitem{eroi}
S.~Moch and A.~Vogt,
Phys.\ Lett.\  B  631 (2005) 48

\bibitem{kt}
J. ~Kodaira and  L.~Trentadue, Phys. Lett. B 112 (1982) 66.

\bibitem{vogt:00} A. Vogt,
Phys. Lett. B 497 (2001) 228.



\bibitem{Aglietti:2002ew}
U.~Aglietti and G.~Ricciardi,
Phys.\ Rev.\ D  66 (2002) 074003

\bibitem{americani} R. Akhouri and I.~Rothstein, Phys. Rev. D {\bf54} (1996) 2349; G.~Korchemsky and G. Sterman, Phys. Lett. B 340 (1994)  96.

\bibitem{future}
 L. Di Giustino,  G. Ricciardi and L. Trentadue, in preparation.



\end{thebibliography}
\end{document}